\documentclass[11pt]{article}
\usepackage{amssymb}
\usepackage{amsmath}
\usepackage{lscape}
\usepackage[dvips]{epsfig}
\def\nuc#1#2{\relax\ifmmode{}^{#1}{\protect\text{#2}}\else${}^{#1}$#2\fi}

\makeatletter
\newcommand{\l@vveden}[2]{\hbox to\textwidth{{\bf \quad #1 #2}}}

\makeatother

\begin{document}

{\bf \Large 
\noindent{
LANL Report LA-UR-00-3600, Los Alamos (2000)\\
}}

{\bf \Large 
\noindent{
Proc. SATIF5, July 18-21, 2000, Paris, France}}


\begin{center}
{ \bf
STUDY OF RESIDUAL PRODUCT NUCLIDE YIELDS FROM 0.1, 0.2, 0.8, 
AND 2.6 GEV PROTON-IRRADIATED \nuc{nat}{Hg} TARGETS}

\vspace{1.2cm}

{\bf Yury E.~Titarenko, Oleg V.~Shvedov, Vyacheslav F.~Batyaev, 
Valery M.~Zhivun, \\
Evgeny I.~Karpikhin, Ruslan D.~Mulambetov, Dmitry V.~Fischenko, \\ 
Svetlana V.~Kvasova}\\
Institute for Theoretical and Experimental Physics,
B. Cheremushkinskaya 25, 117259 Moscow, Russia,

\vspace{0.1cm}
{\bf Stepan G.~Mashnik, Richard E.~Prael, Arnold J.~Sierk} \\
Los Alamos National Laboratory, Los Alamos, NM 87545, USA  \\

\vspace{0.3cm}
{\bf Hideshi Yasuda} \\
Japan Atomic Energy Research Institute, Tokai, Ibaraki, 319-1195, Japan \\

\vspace{1.2cm}

\end{center}

\section*{Abstract}

The direct $\gamma$-spectrometry method is used to measure more than 
350 residual product nuclide 
yields from 0.1, 0.2, 0.8, and 2.6 GeV proton irradiated \nuc{nat}{HgO} 
targets. The $\gamma$-spectrometer resolution 
is of 1.8 keV at the 1332 keV $\gamma$-line. The resultant 
$\gamma$-spectra are 
processed by the GENIE2000 code. The $\gamma$-lines 
are identified, and the cross-sections calculated, by the ITEP-designed 
SIGMA code using the 
PCNUDAT radioactive database. The \nuc{27}{Al}(p,x)\nuc{22}{Na} reaction 
is used as monitor. The experimental results 
are compared with calculations by the LAHET, CEM95, CEM2k, INUCL, CASCADE, 
and YIELDX codes.

\section*{Introduction}

Mercury is planned to be used as a target material in all the present-day 
designs of the Spallation Neutron Source (SNS) facilities 
\cite{bib1}--\cite{bib3}, thus necessitating that the 
proton-Hg interaction characteristics should be studied in a broad 
energy range from a few MeV to 2-3 GeV. Among the characteristics, 
the yields of residual 
product nuclei are of particular importance. They are independent 
nuclear constants to be used 
in practical calculations as well as to verify the codes for calculating 
the SNS facility design parameters, 
such as radioactivity (both current and residual), deterioration of 
resistance to corrosion, yields of 
gaseous products, poisoning, etc.

\section*{Experiment}
 
The experimental samples are 10.5-mm diameter discs manufactured 
by pressing fine-dispersed \nuc{nat}{HgO} powder. The weight contents 
of impurities in the samples do not exceed 0.16\%, of 
which 0.01\% Si, 0.03\% Cl, 0.02\% Ca, 0.04\% Ti, 0.03\% Fe, and 0.01\% Ba. 
The total content of the rest 60 
elements, found by the spark mass-spectrometry, is below 0.02\%.

The measurements were made by the relative method,
 using the \nuc{27}{Al}(p,x)\nuc{22}{Na} reaction to monitor the 
process. The monitors are 10.5-mm Al foils with chemical impurities 
below 0.001\%. Two 
independent proton beams from the ITEP U-2 synchrotron are used to 
irradiate the samples, namely, the 
low-energy (70-200 MeV) and high-energy (800-2600 MeV) beams.

Table \ref{tab1} presents the characteristics of the experimental samples 
and monitors together with the 
main irradiation parameters.

The techniques for irradiating the samples and for processing the 
$\gamma$-spectra are presented in \cite{bib4} 
together with formulas used to determine the fragment nuclide yields.

\begin{table}  
\vspace*{-0.5cm}\begin{center}
\caption{Parameters of experimental samples and monitors, 
irradiation conditions, and 
monitor reaction cross sections} \vspace{2pt}   \label{tab1}
\vspace*{0.5cm}
\begin{tabular}{|c|c|c|c|c|c|}\hline
Proton &Sample &Monitor &Irradiation &Mean proton& \nuc{27}{Al}(p,x)\nuc{22}{Na}	 \\ 
energy, &thickness, & thickness, 	&time, &flux density,& cross section,	 \\ 
GeV&mg/cm$^2$ & mg/cm$^2$	& min& p/cm$^2$/s&	mb \\ \hline
0.10	&	536.0	&	138.2	&	60	&	2.5$\cdot$10$^9$ & 19.1 $\pm$ 1.3	 \\ \hline
0.20	&	537.4	&	137.3	&	45	&	7.1$\cdot$10$^9$ & 15.1 $\pm$ 0.9	 \\ \hline
0.80	&	529.3	&	139.1	&	15	&	1.5$\cdot$10$^10$ & 15.5 $\pm$ 0.9	 \\ \hline
2.6	&	536.3	&	137.0	&	60	&	4.9$\cdot$10$^10$ & 11.7 $\pm$ 0.9	 \\ \hline
 \end{tabular} \end{center}
\end{table}

\section*{Results}
\noindent

More than 350 yields of residual nuclei (from \nuc{22}{Na}
\footnote{The \nuc{22}{Na} and \nuc{24}{Na} yields have been determined 
disregarding the contributions from the Al monitor samples.}  
to \nuc{203}{Hg}) from 0.1, 0.2, 0.8, and 2.6 GeV 
proton-irradiated \nuc{nat}{Hg} have been measured. 
Figs. 1-9
show the products that were measured at all the 
four energies.

\begin{figure}[t!]
\begin{center}
\includegraphics[width=15cm]{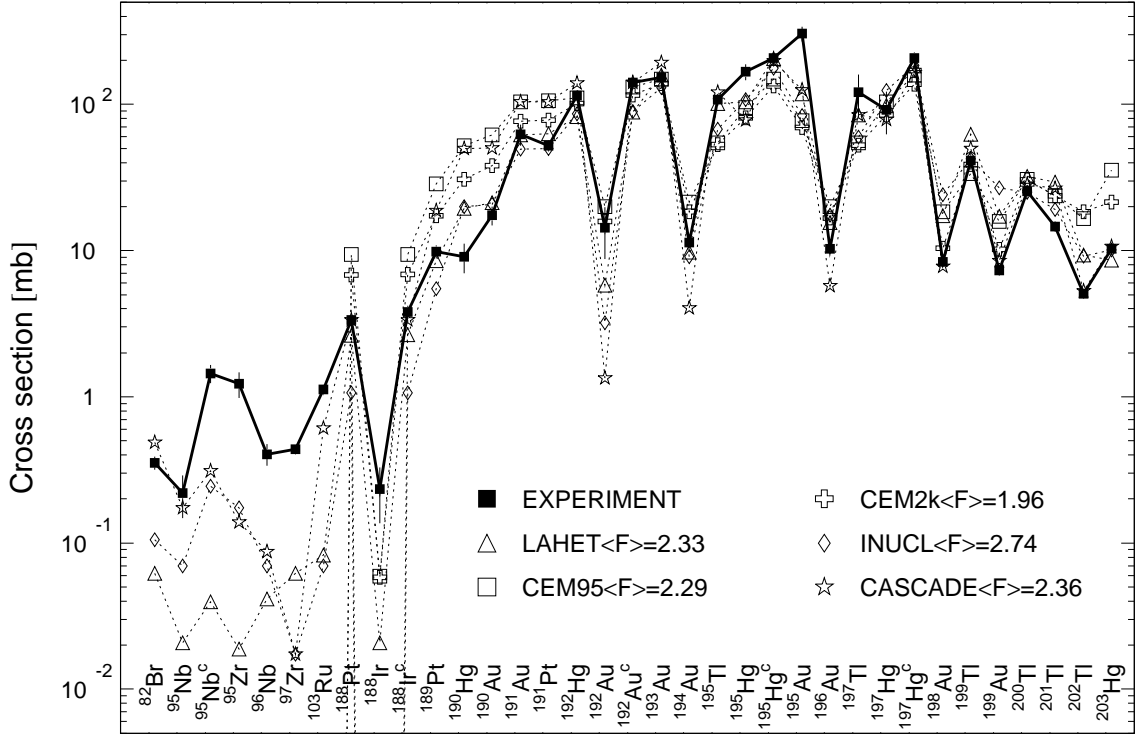}
\caption{Nuclide-by-nuclide comparison between the experimental 
and simulated results for 0.1 GeV 
protons. 
The cumulative yields are labeled with a ``c" when the respective 
independent yields are also shown.}
\label{fig1} 
\end{center} 
\end{figure}

\begin{figure}[t!] 
\begin{center}
\includegraphics[width=15cm]{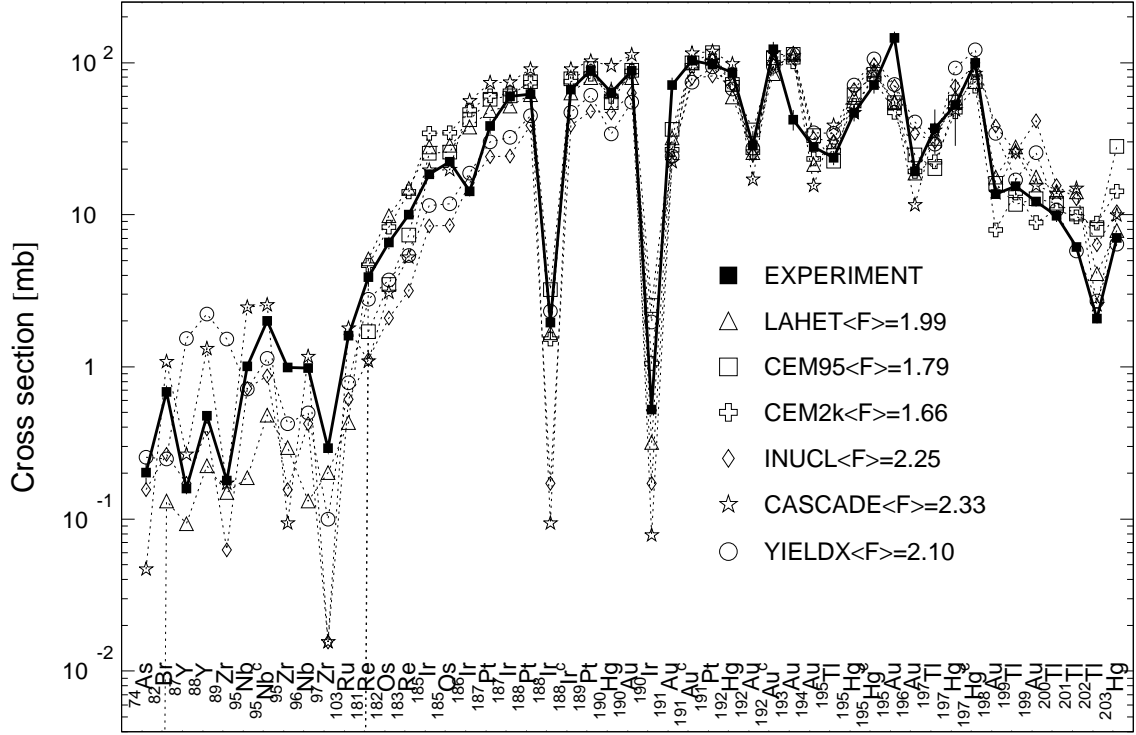}
\caption{Product-by-product comparison between the experimental and 
simulated results for 0.2 GeV 
protons. 
The cumulative yields are labeled with a ``c" when the respective 
independent yields are also shown.}
\label{fig2} 
\end{center} 
\end{figure}

\begin{figure}[t!] 
\begin{center}
\includegraphics[width=15cm]{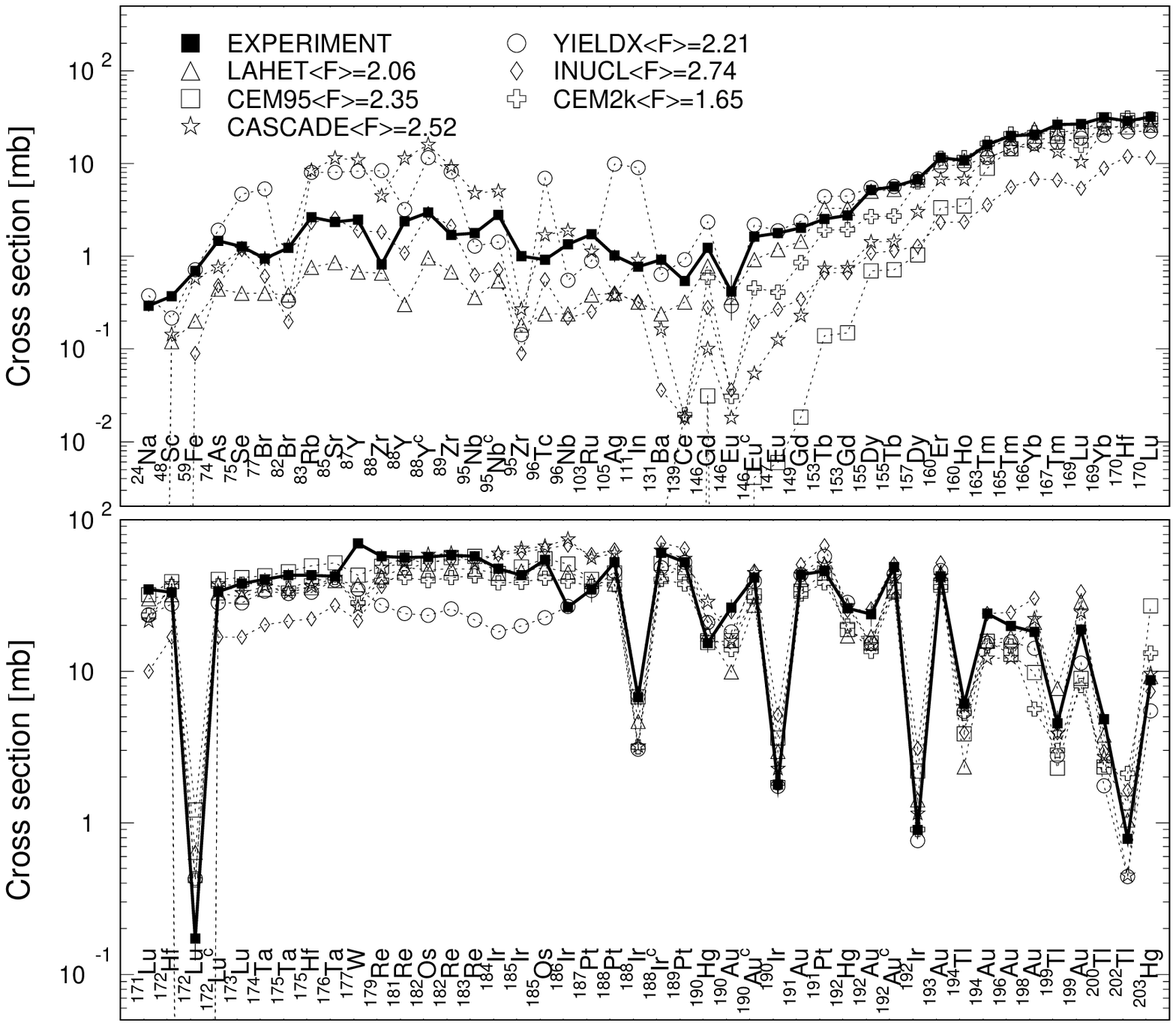}
\caption{Nuclide-by-nuclide comparison between the experimental and 
simulated results for 0.8 GeV 
protons. 
The cumulative yields are labeled with a ``c" when the respective 
independent yields are also shown.}
\label{fig3} 
\end{center} 
\end{figure}

\begin{figure}[t!] 
\begin{center}
\includegraphics[width=15cm]{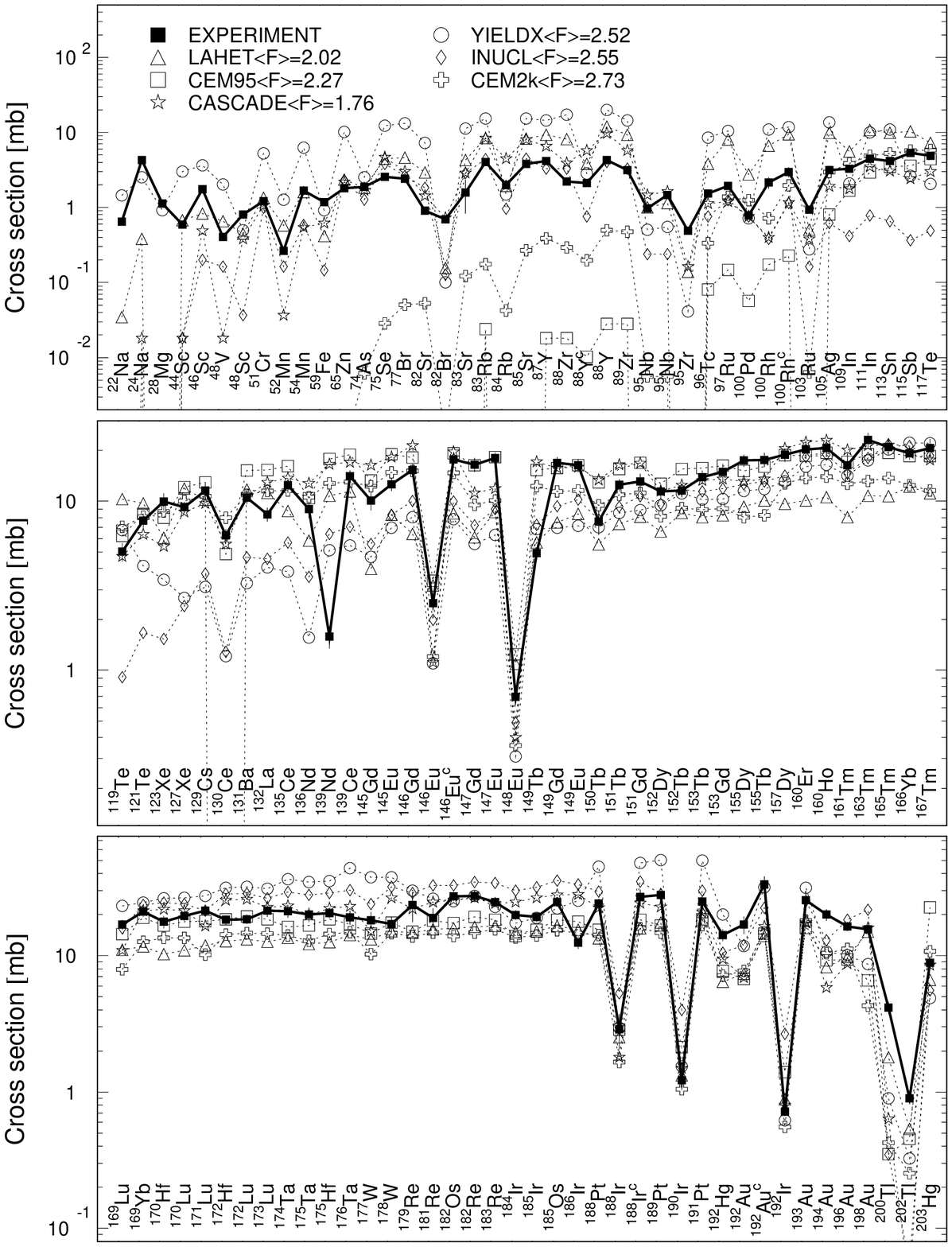}
\caption{Nuclide-by-nuclide comparison between the experimental and 
simulated results for 2.6 GeV 
protons. 
The cumulative yields are labeled with a ``c" when the respective 
independent yields are also shown.}
\label{fig4} 
\end{center} 
\end{figure}

The experimental data were compared with the LAHET, CEM95, CEM2k, INUCL, 
CASCADE, and YIELDX code-simulated yields. The comparison method and a
short description of all codes together with further
references  
may be found in \cite{bib4}. It should be noted that all these codes
do not calculate the independent and cumulative yields individually 
for the ground and 
metastable states of the produced radionuclides, whereas the yields 
of either ground or metastable states 
alone are often measured. Therefore, those particular measurement results 
were excluded from the 
comparison procedure.
The only exclusion is the case of measuring both states, so the total yields 
can be compared with the simulation results. 
Table \ref{tab2} and Figs. \ref{fig1}-\ref{fig4} present the results 
of a detailed
nuclide-by-nuclide comparison.

\begin{table}  
\vspace*{-0.5cm}
\caption{Statistics of the experimental-to-simulated yield comparisons} \vspace{2pt}   \label{tab2}
\vspace*{0.5cm}
\begin{tabular}{|c|c|c|c|c|c|c|}\hline
Code	&\multicolumn{3}{|c|}{E$_p$ = 0.1 GeV, N$_T$ = 48, N$_G$ = 35} 	&\multicolumn{3}{|c|}{E$_p$ = 0.2GeV, N$_T$ = 66, N$_G$ = 49} \\ \cline{2-7}		
	&N$_{C1.3}$ / N$_{C2.0}$ / N$_S$&	$<$F$>$&	S($<$F$>$)&	N$_{C1.3}$ / N$_{C2.0}$ / N$_S$&	$<$F$>$&	S($<$F$>$) \\ \hline
LAHET	&	13/21/30	&	2.24	&	1.92	&	22/36/48	&	1.99	&	1.69	 \\ \hline
CEM95	&	6/15/28	&	2.29	&	1.61	&	20/31/38	&	1.79	&	1.56	 \\ \hline
CEM2k	&	9/18/28	&	1.96	&	1.52	&	22/31/38	&	1.66	&	1.48	 \\ \hline
INUCL	&	10/19/33	&	2.74	&	2.05	&	12/27/48	&	2.25	&	1.66	 \\ \hline
CASCADE	&	16/24/33	&	2.36	&	1.98	&	21/35/48	&	2.33	&	1.95	 \\ \hline
YIELDX	&	--	&	--	&	--	&	14/35/49	&	2.10	&	1.66	 \\ \hline
	&\multicolumn{3}{|c|}{E$_p$ = 0.8GeV, N$_T$ = 106, N$_G$ = 88}	&\multicolumn{3}{|c|}{E$_p$ = 2.6 GeV, N$_T$ = 142, N$_G$ = 121} \\ \cline{2-7}
	&N$_{C1.3}$ / N$_{C2.0}$ / N$_S$&	$<$F$>$&	S($<$F$>$)&	N$_{C1.3}$ / N$_{C2.0}$ / N$_S$&	$<$F$>$&	S($<$F$>$) \\ \hline
LAHET	&	42/63/87	&	2.06	&	1.73	&	23/80/118	&	2.02	&	1.49	 \\ \hline
CEM95	&	30/46/59	&	2.35	&	2.13	&	46/77/91	&	2.27	&	2.11	 \\ \hline
CEM2k	&	26/51/63	&	1.65	&	1.43	&	32/77/101	&	2.73	&	2.29	 \\ \hline
INUCL	&	26/42/82	&	2.74	&	1.95	&	37/77/115	&	2.55	&	2.03	 \\ \hline
CASCADE	&	35/59/84	&	2.52	&	2.10	&	56/93/114	&	1.76	&	1.55	 \\ \hline
YIELDX	&	32/62/88	&	2.21	&	1.82	&	26/65/120	&	2.52	&	1.78	 \\ \hline
 \end{tabular} 
\end{table}

In Table \ref{tab2}, N$_T$ is the total number of the measured yields; 
N$_G$ is the number of the measured yields selected 
to be used in comparison with calculations; 
N$_S$ is the number of the products whose yields were 
simulated by a particular code; 
N$_{C1,3}$ is the number of the comparison events when the 
simulation-experiment difference does not exceed 30\%; 
N$_{C2.0}$ is the number of the comparison events when 
the simulation-experiment difference does not exceed a factor of 2.

\begin{figure}
\begin{center}
\includegraphics[angle=-90, width=15cm]{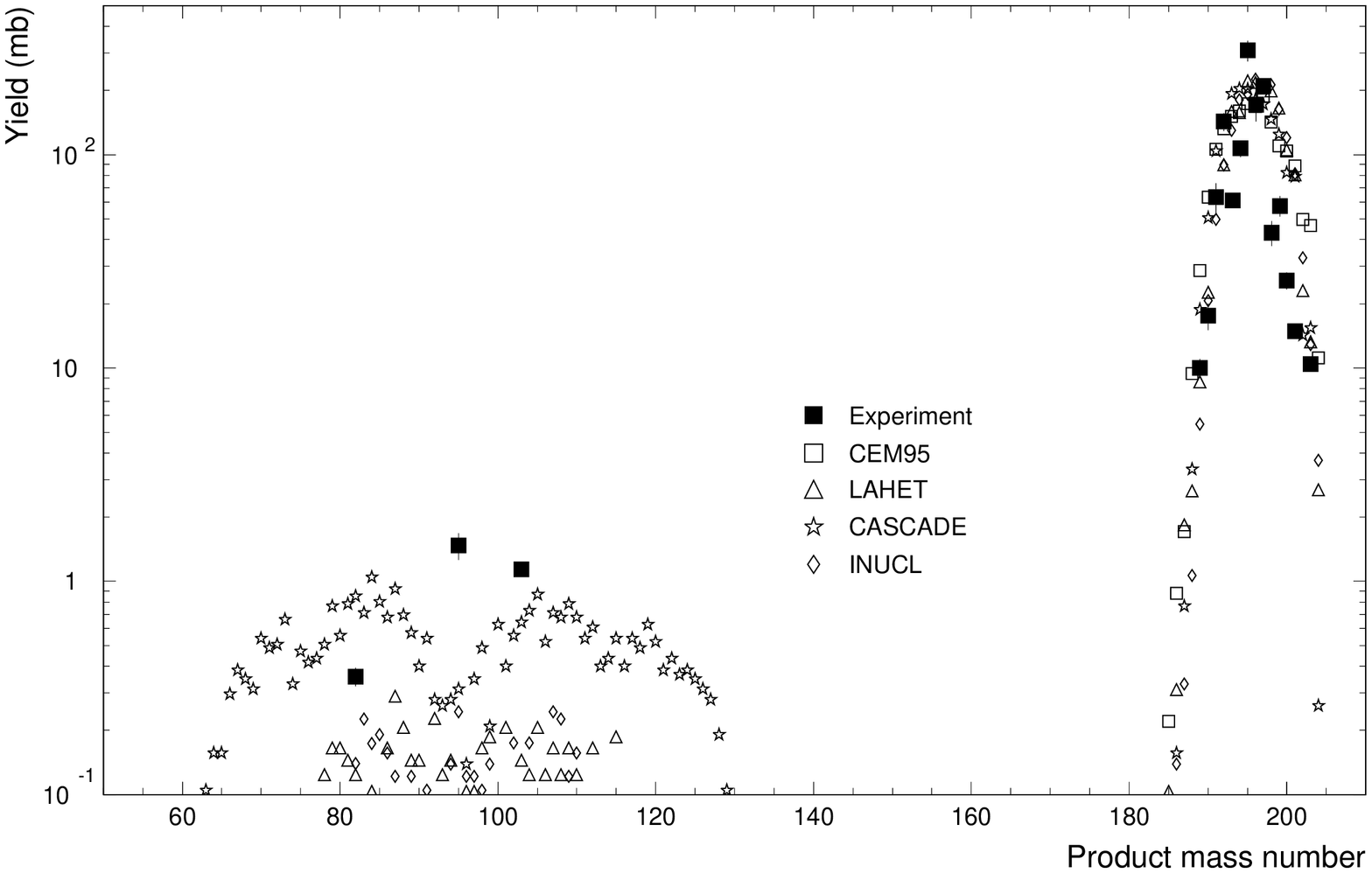}
\caption{The simulated and experimental mass yields at 0.1 GeV.}
\label{fig5} \end{center} \end{figure}
 
\begin{figure} 
\begin{center}
\includegraphics[angle=-90, width=15cm]{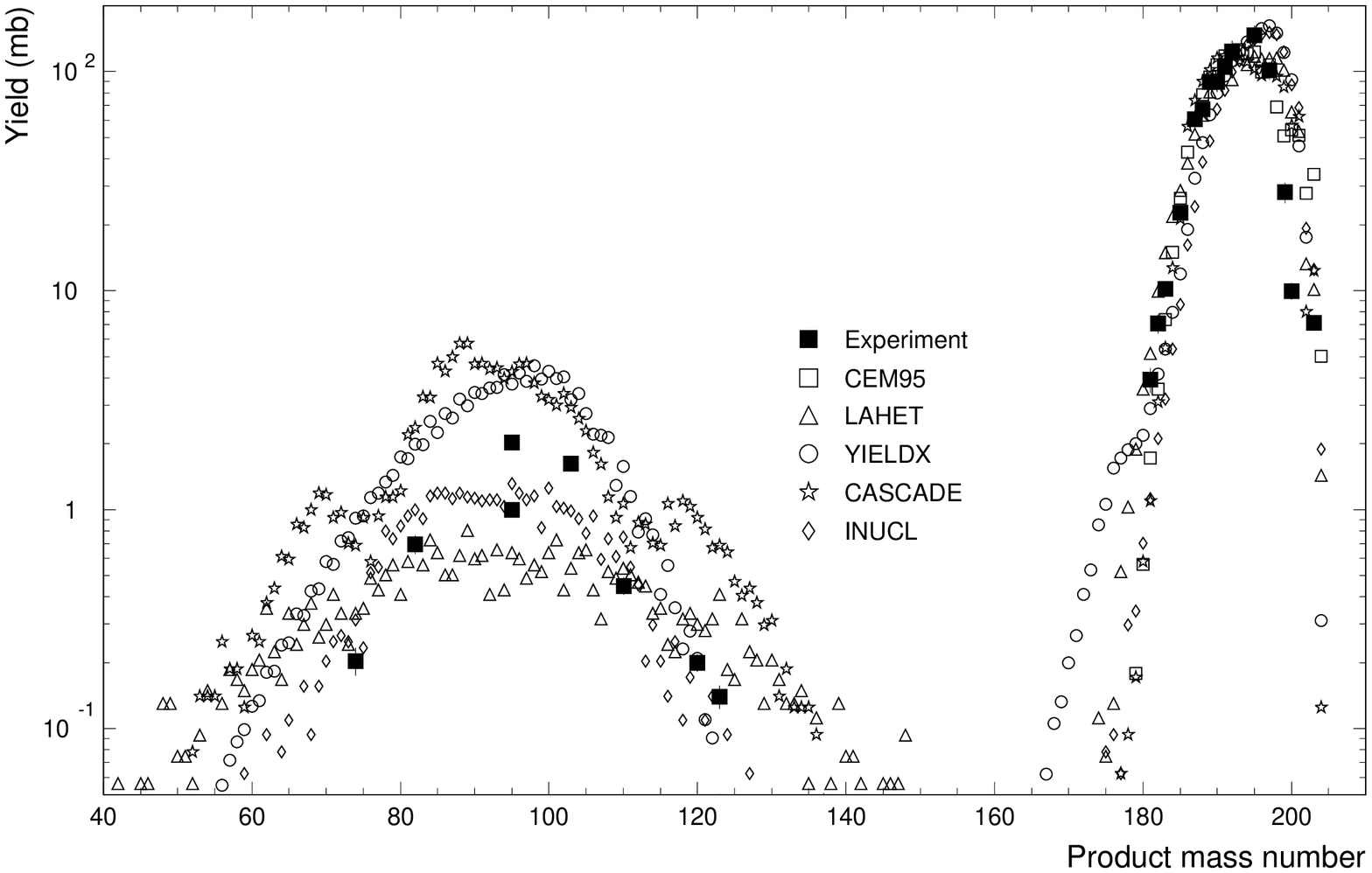}
\caption{The simulated and experimental mass yields at 0.2 GeV.}
\label{fig6} \end{center} \end{figure}
 
\begin{figure} 
\begin{center}
\includegraphics[angle=-90, width=15cm]{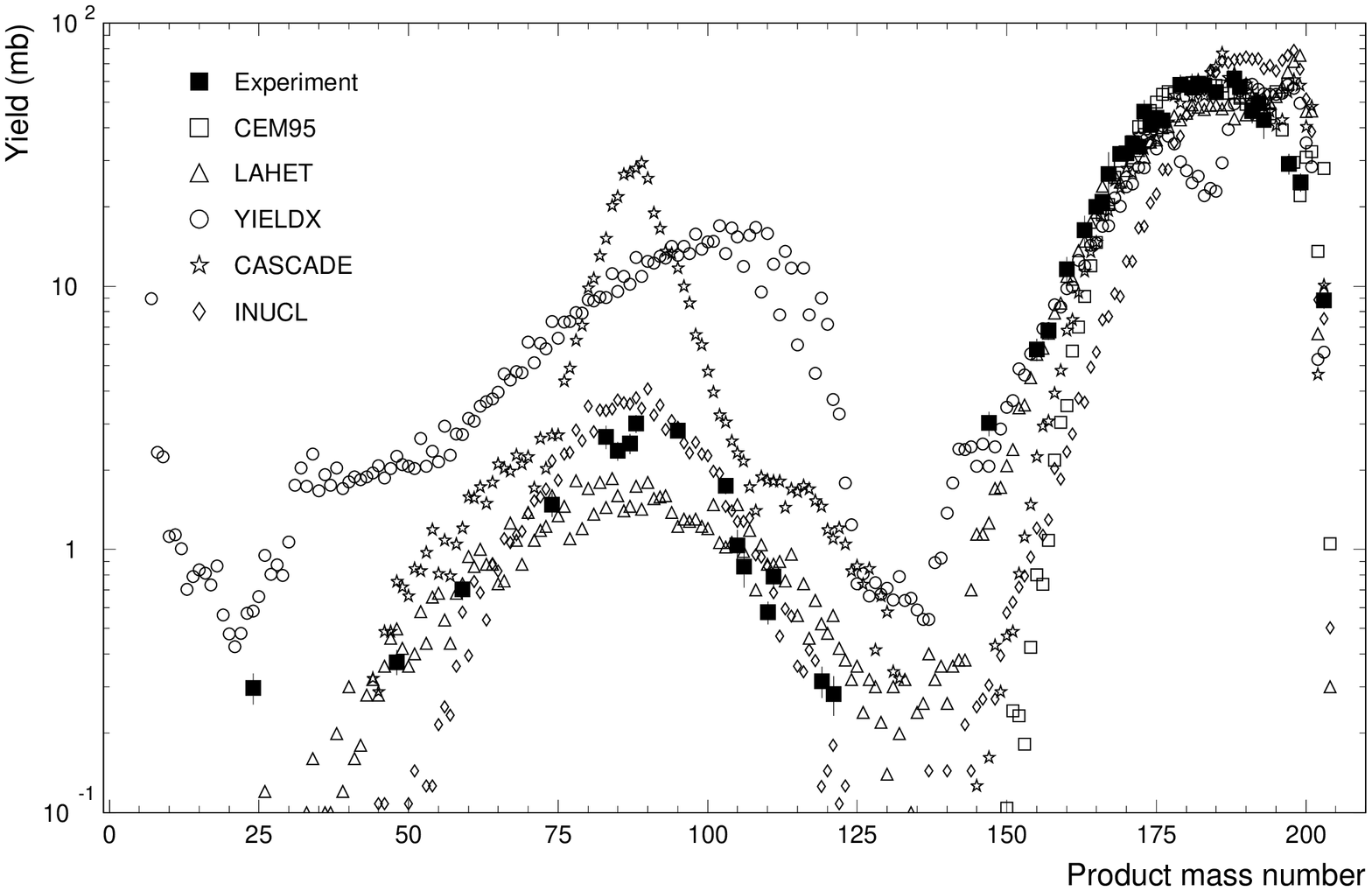}
\caption{The simulated and experimental mass yields at 0.8 GeV.}
\label{fig7} \end{center} \end{figure}
 
\begin{figure} 
\begin{center}
\includegraphics[angle=-90, width=15cm]{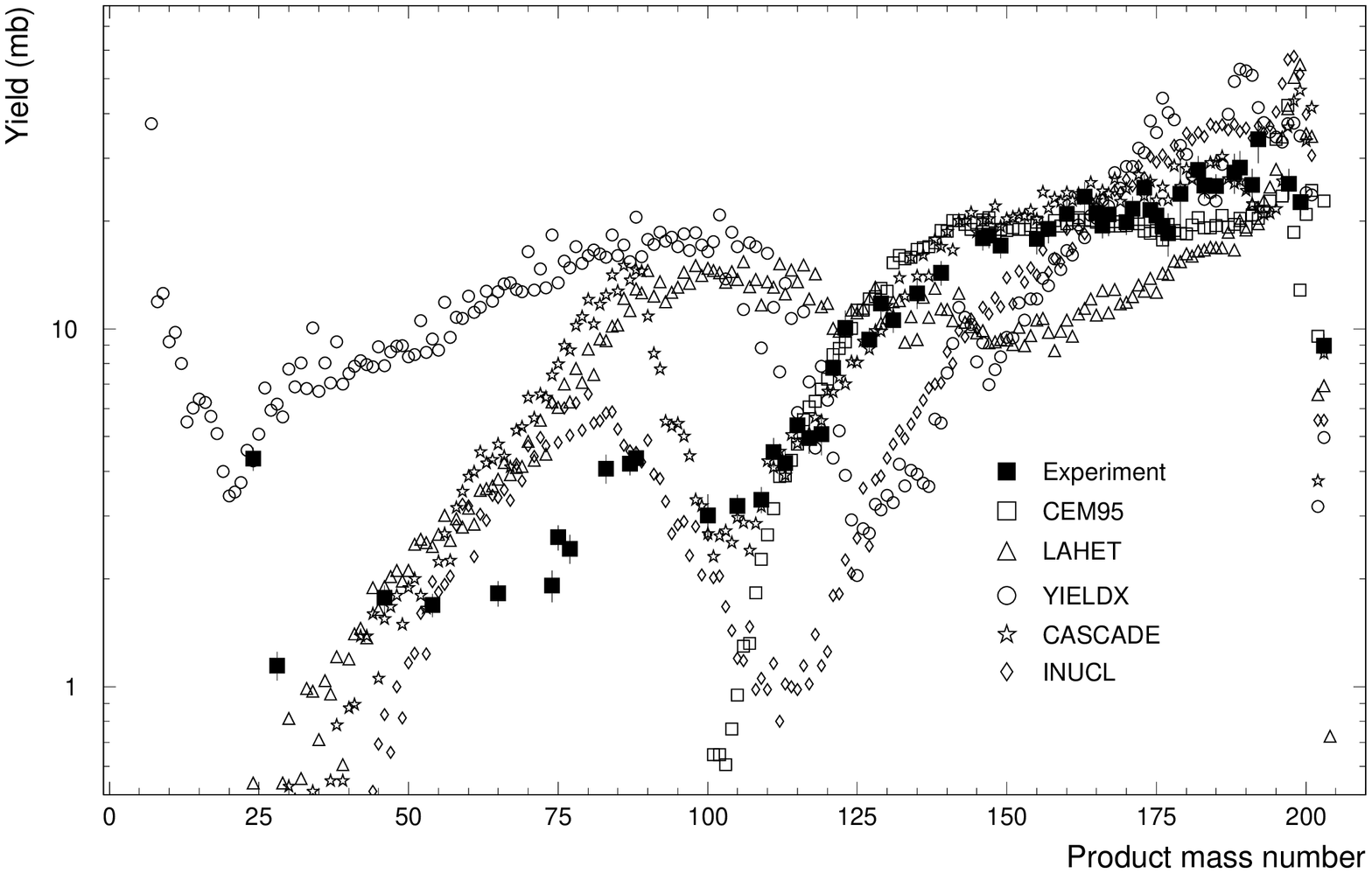}
\caption{The simulated and experimental mass yields at 2.6 GeV.}
\label{fig8} \end{center} \end{figure}
 
\begin{figure} 
\begin{center}
\includegraphics[width=14cm]{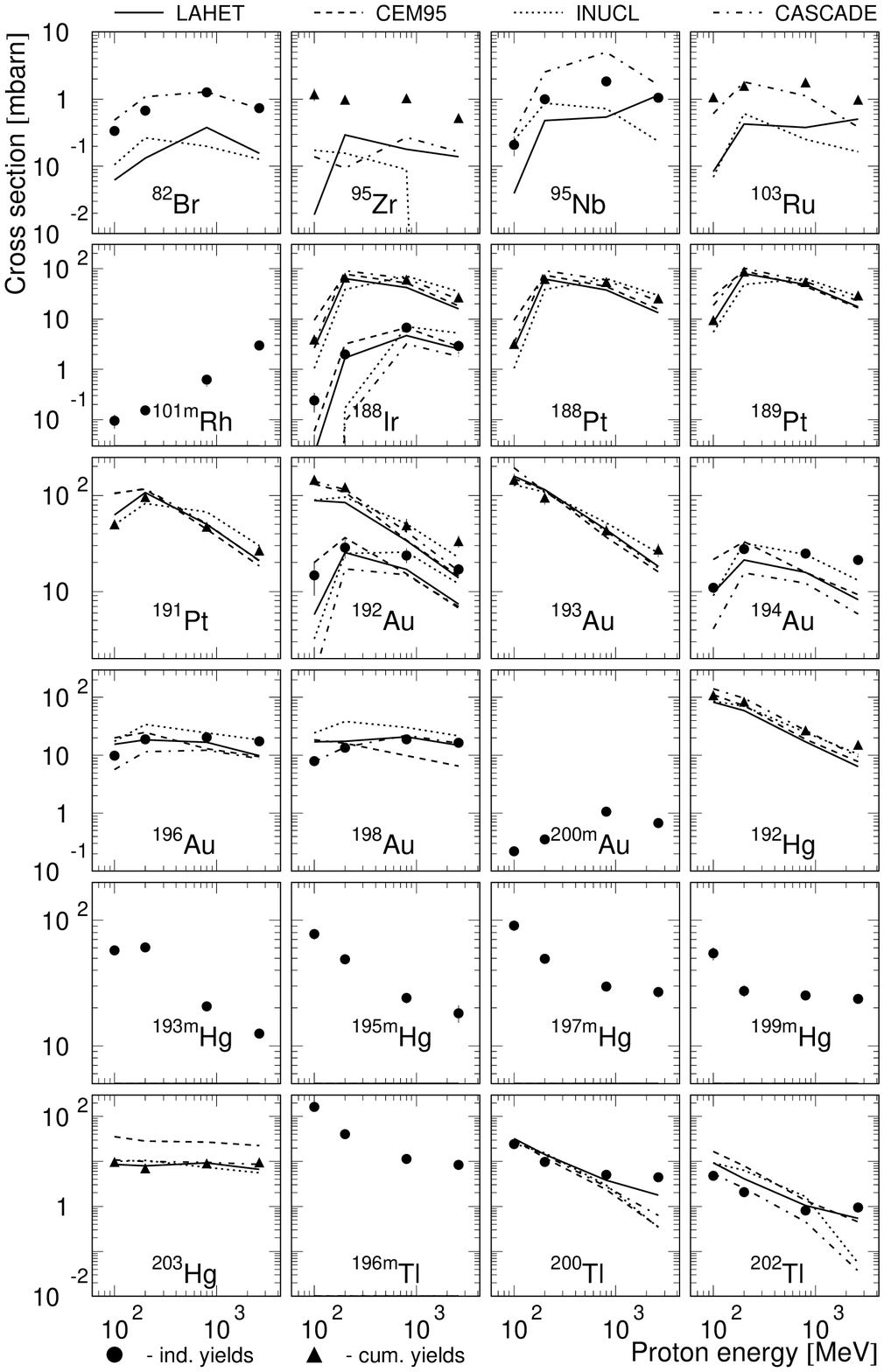}
\caption{Some experimental and simulated yields versus the proton energy.}
\label{fig9} \end{center} \end{figure}

Figs. \ref{fig1}-\ref{fig4} show the results of the nuclide-by-nuclide 
comparison of the experimental data with 
the LAHET, CEM95, CEM2k, CASCADE, INUCL, and the YIELDX code-simulated 
results. 
One can see
that all codes (except YIELDX) adequately predict 
the A$>$170 product yields for 100, 
200, and 800 MeV protons and the A$>$120 product yields for 2600 MeV protons. 
The yields simulated 
by all the codes in the remaining ranges of masses,
i.e., in the fission and fragmentation regions, 
are very different from experiment, with 
the most significant differences observed in the 80$<$A$<$103 range for 
100 and 200 MeV protons, in the 48$<$A$<$130 range for 800 MeV protons, 
and in the 28$<$A$<$100 range for 2600 MeV protons.
It should be noted that 
CEM95 and CEM2k
do not calculate the process of fission itself, and do
not provide fission fragments and a further possible evaporation of
particles from them. When, during a Monte Carlo simulation of a
compound stage of a reaction using the evaporation and fission widths  
these codes have to simulate a fission, 
they simply remember this event (that 
permits them to calculate fission cross section and fissility)
and finish the calculation of this event 
without a real subsequent calculation of fission fragments.
Therefore, the results from 
CEM95 and CEM2k shown here
reflect the contribution to
the total yields of the nuclides only from deep spallation processes of 
successive emission of particles from the target, 
but do not contain fission products. 
To be able to describe 
nuclide production in the fission region, these codes have to be extended
by incorporating a model of high energy fission
(e.g., in the transport code MCNPX, where these code are used, 
they are complemented 
by the RAL fission model).

Figs. \ref{fig5}-\ref{fig8} show the simulated mass yield of the 
reaction products. 
The experimental cumulative yields, which are often equal to the mass 
yields within measurement errors,
are also shown for comparison.
The following conclusions may be drawn from the comparison between 
the experimental and 
simulated mass yields:
\begin{itemize}
\item in the case of 0.1 GeV protons and A$>$190, 
all codes predict the mass curve shape quite adequately,
 \item in the case of 0.2 GeV protons and A$>$180, 
 all code-simulated yields are in a good agreement with the data,
 \item in the case of 0.8 GeV protons, the best agreement with 
 experiment is obtained by YIELDX, for A$>$130 and by INUCL, for A$<$130,
 \item in the case of 2.6 GeV protons and A$>$100, 
 the CEM95 and CASCADE results
agree with the data, while the LAHET calculations are 
underestimated and the YIELDX and 
INUCL yields represent the mass curve shape erroneously. 
None of the codes 
can describe well the experimental curve shape at A$<$100.
\end{itemize}

Fig. \ref{fig9} illustrates the dependence of a part of measured yields 
on the proton energy, i.e., excitation functions. These data 
will be analyzed further
after a final release of all the experimental results.

\section*{Conclusion}
\noindent

Our experiment-to-simulation comparison has shown that, on the whole, 
the simulation codes can 
but poorly predict the experimental results. In some cases, the 
differences reach an order of magnitude or even more. 
The last version of the improved cascade-exciton model code, CEM2k [5],
shows the best agreement with the data in the spallation region at all 
energies except 2.6 GeV, where the model overestimates the expected
experimental fission cross section of about 75 mb [6] by a factor of 4.
This overestimation of the fission cross section causes an underestimation
of the yield of nuclei which are most likely to fission at the evaporation
stage of a reaction, after the cascade and preequilibrium stages , i.e.,
for 170 $<$ A $<$ 185. (Similar disagreement with the data one can see as
well for LAHET and CEM95, that is also related with an overestimation
of the fission cross section at 2.6 GeV). The code CEM2k is still under
development, its problem with the overestimation of fission cross
sections at energies above 1 GeV has yet to be solved, and it has to be 
complemented with a model of fission fragment production as mentioned
above, to be able to describe as well fission products.

This means that the nuclear data must be accumulated persistently in the 
above-mentioned ranges of energies and masses,
to help improve hadron-nucleus interaction models released
in codes used in applications.

\section*{Acknowledgement}

The work was carried out under the ISTC Project \#839 supported by 
the European Community, Japan (JAERI), 
Norway and, partially, by the U. S. Department of Energy.

\end{document}